\documentclass[prd,aps,preprint,tightenlines,nofootinbib,superscriptaddress]{revtex4}
\usepackage{epsfig}
\usepackage{amsmath}
\input{epsf}
\usepackage{psfrag}

\usepackage[usenames,dvipsnames]{color}

\begin{document}


\vspace*{2cm}
\title{Coulomb nuclear interference effect in dipion production in ultraperipheral heavy ion collisions}

\author{Yoshikazu Hagiwara}
 \affiliation{\normalsize\it Key Laboratory of
Particle Physics and Particle Irradiation (MOE),Institute of
Frontier and Interdisciplinary Science, Shandong University,
(QingDao), Shandong 266237, China }

\author{Cheng~Zhang}
 \affiliation{\normalsize\it Key Laboratory of
Particle Physics and Particle Irradiation (MOE),Institute of
Frontier and Interdisciplinary Science, Shandong University,
(QingDao), Shandong 266237, China }

\author{Jian~Zhou}
 \affiliation{\normalsize\it Key Laboratory of
Particle Physics and Particle Irradiation (MOE),Institute of
Frontier and Interdisciplinary Science, Shandong University,
(QingDao), Shandong 266237, China }

\author{Ya-jin Zhou}
\affiliation{\normalsize\it Key Laboratory of Particle Physics and
Particle Irradiation (MOE),Institute of Frontier and
Interdisciplinary Science, Shandong University, (QingDao), Shandong
266237, China }

\begin{abstract}
We study exclusive $\pi^+ \pi^-$ pair production near $\rho^0$ resonance peak in ultraperipheral heavy ion collisions.  Pion pair can either be produced  via two photon fusion process or  from the decay of  $\rho^0$ in photon-nuclear reaction.   At very low pair transverse momentum,  the electromagnetic and nuclear amplitudes become comparable. We show that the Coulomb nuclear interference amplitude   gives rise to sizable   $\cos \phi$ and $\cos 3\phi$ azimuthal asymmetries, which can be used to constrain the phase of the dipole-nucleus scattering amplitude. 
\end{abstract}

\maketitle

Strong electromagnetic fields induced by relativistic heavy ions can be effectively viewed as a flux of quasi-real photons, which have been used to study a wide variety of physics in ultraperipheral heavy ion collisions(UPCs). Among many interesting UPC related topics, the diffractive   vector mesons photoproduction on nuclei is one of the main focus of UPC physics as it offers  access to the  gluon tomography of nucleus as well as the tests of the CGC description of saturation physics~\cite{Ryskin:1992ui,Brodsky:1994kf,Klein:1999qj,Munier:2001nr, Kowalski:2003hm,Kowalski:2006hc,Strikman:2008zz,Rebyakova:2011vf,Guzey:2013qza,Lappi:2010dd,Xie:2016ino,Cai:2020exu,Hagiwara:2017fye,Hatta:2019ocp}. There are also many active experimental programs  devoted to the investigations of  diffractive vector meson production~\cite{Khachatryan:2016qhq,Adamczyk:2017vfu,STAR:2019yox,Sirunyan:2019nog,Acharya:2020sbc}  at both RHIC and LHC recently.

 Experimentally, the vector meson $\rho^0$ events are  reconstructed by measuring  its decay product   $\pi^+ \pi^-$ pair. The invariant mass spectrum of dipion can be well fitted with a relativistic Breit-Wigner resonance for $\rho^0$ plus a flat direct pion pair continuum.   Such continuum background receives contributions from several different sources~\cite{Soding:1965nh,Brodsky:1971ud,Klusek-Gawenda:2013rtu,Bolz:2014mya}: e.g. photon-pomeron fusion process, and direct dipion  production in photon-photon collisions.  Since  quasi-real photons are almost on shell, pion can be approximately treated as a point like scalar particle in the low invariant mass region of dipion system.  As a result, the amplitude of  $\gamma \gamma \rightarrow \pi^+ \pi^-$ process can be straightforwardly  computed using the scalar QED. The amplitude of the electromagnetic production is much smaller than that of the photonuclear interactions in  most kinematic regions. However, this is not always the case, as the long range of the electromagnetic forces leads to rapidly growing amplitude with decreasing nuclear recoil momentum squared $t$, while the nuclear amplitude approaches a constant with vanishing $t$.  At sufficiently low pair transverse momentum($q_\perp$),  the pure electromagnetic and the nuclear amplitudes become  comparable, and in particular, the Coulomb-nuclear interference(CNI)  amplitude could be sizable. 

 Quite interestingly, we found that such CNI effect does not contribute to the azimuthal averaged cross section of the exclusive dipion production.  Instead, it contributes to  the  azimuthal asymmetric cross section.  To be more specific, the linear polarization of quasi-real photons can induce distinctive $\cos \phi$ and $\cos 3\phi$  modulations, where $\phi$ is defined as the azimuthal angle between $q_\perp$ and pion's transverse momentum $P_\perp$.   The sign and the magnitude of these azimuthal  asymmetries are sensitive to the relative phase between the EM amplitude and the QCD amplitude.  These observables thus can  be used to  constrain the phase of photonuclear amplitude in future experiments. 

The key ingredient of our analysis presented in this paper is the  linear polarization of  quasi-real photons. This phenomenon was not recognized until very recently~\cite{Li:2019yzy,Li:2019sin}.  The linear polarization of coherent photons can give rise to a characteristic $\cos 4\phi$ azimuthal asymmetry in pure EM dilepton production in UPCs~\cite{Li:2019yzy,Li:2019sin,Xiao:2020ddm}, which later was clearly seen in the STAR measurement~\cite{Adam:2019mby}.  The computed impact parameter  dependent asymmetry turns out to be in excellent agreement with the experimental data for the UPC case.  With it being experimentally  confirmed, the linearly polarized quasi-real photon beam in heavy ion collisions can be used as a powerful tool to explore the novel QCD phenomenology. For example,   the significant $\cos 2\phi$ and $\cos 4\phi$ asymmetries for  $\rho^0$ meson production in UPCs have been reported by STAR collaboration previously~\cite{daniel:2019}. The observed asymmetries result from incident photon's linear polarization.  Its unique diffractive pattern crucially depends on transverse  spacial distribution of the gluonic matter inside nucleus, and   double slit like quantum interference effect~\cite{Xing:2020hwh}.

The paper is organized as follows. In Sec.II, we derive all cross section formulas including pure EM dipion production contribution, these from the decay of $\rho^0$ meson and their interference terms. In Sec.III, we present the numerical estimations of $\cos  \phi$ and  $\cos  3\phi$ azimuthal asymmetries for exclusive dipion production.  The paper is summarized in Sec. IV.

 \section{The Coulomb nuclear interference effect}
The vector meson photoproduction  is conventionally computed using  the dipole model~\cite{Ryskin:1992ui,Brodsky:1994kf}, in which the whole process is divided into  three steps: quasi-real photon splitting into quark and anti-quark pair, the color dipole scattering off nucleus, and subsequently recombining  to form  a vector meson after penetrating the nucleus target. Following this picture, one can directly write down the polarization averaged cross section of the vector meson production in ep or eA collisions. For the photoproduction of vector meson in UPCs, it is important to take into account double slit like quantum interference effect~\cite{Klein:1999gv,Abelev:2008ew,Zha:2018jin}.  To this end, we developed a joint impact parameter dependent and $q_\perp$ dependent cross section formula in a previous work.  The resulting unpolarized  cross section gives excellent description to the STAR experimental data.  We further computed the $\cos 2\phi$ azimuthal asymmetry induced by the linear polarization of photons, and are able to  describe its $q_\perp$ dependent behavior reasonably well~\cite{Xing:2020hwh}.

One  can easily derive the dipion production amplitude  by multiplying the $\rho^0$ production amplitude with  a simplified Breit-Wigner form which describes   the transition from $\rho^0$ to $\pi^+ \ \pi^-$,
\begin{eqnarray}
{\cal M}_{\rho \rightarrow \pi^+ \pi^-}={\cal M}_{\rho} f_{\rho \pi \pi}
 \frac{ P_\perp \cdot \epsilon_\perp^{V}}{Q^2-M_\rho^2+iM_\rho \Gamma_\rho}
\end{eqnarray}
where ${\cal M}_{\rho} $ denotes  $\rho^0$ production amplitude. $ \epsilon_\perp^{V}$ and $M_\rho$ are  $\rho^0$'s polarization vector and mass respectively.  $Q$ is the invariant mass of dipion system and $P_\perp$ is  defined as $P_\perp=(p_{1\perp}-p_{2\perp})/2$ with $p_{1\perp}$ and $p_{2\perp}$ being the produced pions' transverse momenta. $f_{\rho \pi \pi}$ is the effective coupling constant and fixed to be $f_{\rho \pi \pi}=12.24$ according to the optical theorem with the parameter $\Gamma_\rho=0.156 \ {\text {GeV}}$.  By recycling the result from our earlier work~\cite{Xing:2020hwh}, it is straightforward to write down the differential cross section for dipion production,
\begin{eqnarray}
  && \!\!\!\!\!\!\!\!\!\!\!\!
   \frac{d \sigma_{\rho \rightarrow \pi \pi}}{d^2 p_{1\perp} d^2 p_{2\perp} dy_1 dy_2 d^2 \tilde b_{\perp} }  =\frac{1 }{2(2\pi)^7}  \frac{P_\perp^2}{(Q^2-M_\rho^2)^2+M_\rho^2 \Gamma_\rho^2}  f_{\rho \pi \pi}^2 
    \nonumber \\ &&\!\!\!\!\!\!\!\!\times
    \int d^2 \Delta_\perp d^2k_\perp d^2 k_\perp'
    \delta^2(k_\perp+\Delta_\perp-q_\perp)
     (\hat P_\perp \! \cdot \hat k_\perp )(\hat P_\perp\! \cdot \hat k_\perp' )
     \nonumber \\ && \!\!\!\!\!\!\!\!\times \!
    \left \{ \int \!\! d^2   b_\perp e^{i \tilde b_\perp \cdot
    (k_\perp'\!\!-k_\perp)} \left [ T_A(b_\perp) {\cal
    A}_{in}(x_2,\Delta_\perp) {\cal A}_{in}^*(x_2,\Delta_\perp') {\cal
    F}(x_1,k_\perp){\cal F}(x_1,k_\perp')\!
      + \!( A \!\leftrightarrow \! B)
    \right ] \right .\
     \nonumber \\ &&
    + \!\!\left [  e^{i \tilde b_\perp \cdot (k_\perp'\!\!-k_\perp)}
    {\cal A}_{co}(x_2,\Delta_\perp) {\cal A}_{co}^*(x_2, \Delta_\perp')
    {\cal F}(x_1,k_\perp){\cal F}(x_1,k_\perp')
     \right ]
      \nonumber \\ &&
    + \!\!\left [  e^{i \tilde b_\perp \cdot (\Delta_\perp'\!\!-\Delta_\perp)}
    {\cal A}_{co}(x_1,\Delta_\perp) {\cal A}_{co}^*(x_1, \Delta_\perp')
    {\cal F}(x_2,k_\perp){\cal F}(x_2,k_\perp')
     \right ]
    \nonumber \\  &&+ \!\!
      \left [ e^{i \tilde b_\perp \cdot (\Delta_\perp'\!-k_\perp)}
     {\cal A}_{co}(x_2,\Delta_\perp) {\cal A}_{co}^*(x_1, \Delta_\perp'){\cal F}(x_1,k_\perp){\cal F}(x_2,k_\perp')
     \right ]
        \nonumber \\  &&+ \!\!\!\!
     \left .\
     \left [ e^{i \tilde b_\perp \cdot (k_\perp'\!-\Delta_\perp)}
     {\cal A}_{co}(x_1,\Delta_\perp) {\cal A}_{co}^*(x_2, \Delta_\perp'){\cal F}(x_2,k_\perp){\cal F}(x_1,k_\perp')
     \right ]
      \right \}\!
       \label{fcs}
     \end{eqnarray}
  where  $y_1$ and $y_2$ are final state pions'  rapidities. $k_\perp$, $\Delta_\perp$, $k_\perp'$ and $\Delta_\perp'$ are incoming photon's transverse momenta and nucleus recoil transverse momenta in the amplitude and the conjugate amplitude respectively.  $\tilde b_\perp$ is the transverse distance between the center of the two colliding nuclei. The unit transverse vectors are defined following  the pattern as  $\hat k_\perp=k_\perp/|k_\perp|$ and $\hat P_\perp=P_\perp/|P_\perp|$.  ${\cal F}(x_1,k_\perp)$ describes  quasi-real photon distribution amplitude, and will be specified later.
  ${\cal A}_{co}(\Delta_\perp)$ and ${\cal A}_{in}(\Delta_\perp)$ are given by,
     \begin{eqnarray}
      {\cal A}_{co}(x_g,\Delta_\perp) \!\!&=&\!\! \int \!d^2
     b_\perp e^{-i \Delta_\perp \cdot b_\perp} \!\int \frac{d^2
     r_\perp}{4\pi}  \ N(r_\perp,b_\perp) [\Phi^*\!K](r_\perp)
      \nonumber \\
      {\cal A}_{in}(x_g,\Delta_\perp) \!\!&=&\!\! \sqrt{\!A }2 \pi B_p e^{-\!B_p\Delta_{\! \perp}^2\!/2}
         \left [  \int \!\frac{d^2
     r_\perp}{4\pi}  {\cal N}(r_\perp)e^{\!-2\pi (A\!-\!1)B_p T_A(b_\perp) {\cal N}(r_\perp) }
     [\Phi^*\!K](r_\perp) \right ] \ \ \
     \end{eqnarray}
which represent the coherent and incoherent vector meson production amplitudes respectively. 
  ${\cal N}(r_\perp)$ is the dipole-nucleon scattering amplitude. $N(r_\perp,b_\perp)$ is the elementary amplitude for the scattering of a dipole of size $r_\perp$ on a target nucleus at the impact parameter $b_\perp$ of the photon-nucleus collision. $T_A(b_\perp)$ is the nuclear thickness function. The longitudinal momentum fraction transferred to  the vector meson via the dipole-nucleus interaction is given by $x_g= \sqrt{\frac{P_\perp^2+m_\pi^2}{S}}(e^{-y_1}+e^{-y_2})$. And $[\Phi^*\!K]$ denotes the overlap of virtual photon wave function and the vector meson wave function,
     \begin{eqnarray}
     [\Phi^*\!K](r_{\! \perp})\! =\!\frac{N_{\!c} e e_q}{\pi}\!\! \int_0^1 \!\!\! dz\!
     \left \{m_q^2 \Phi^*(|r_\perp|,z)K_0(|r_\perp| e_f) \!+\!
     \left [  z^2\!+\!(1\!-\!z)^2   \right ]
      \!  \frac{\partial\Phi^*(|r_\perp|,z)}{\partial |r_\perp|}
      \frac{\partial  K_0(|r_\perp| e_f)}{\partial |r_\perp|} \!\right \}
     \end{eqnarray}
 where $z$ stands for the  fraction of photon's light-cone momentum carried by quark.
The physical meanings of all other parameters and shorthand notations appear in the above equations can be found in Ref.~\cite{Xing:2020hwh}.

We now proceed to compute pion pair production amplitude in two photon collisions using an effective lagrangian which is the simplest possible form satisfying the gauge invariance,
\begin{eqnarray}
{\cal L}_I=ie F_\pi(Q_\gamma^2) A^\mu \left [(\partial_\mu \pi^\dag) \pi-\pi^\dag \partial_\mu \pi \right ]
-e^2 F_\pi(Q_\gamma^2) A_\mu A^\mu \pi^\dag \pi
\end{eqnarray}
where the EM form factor $F_\pi(Q_\gamma^2)$ can be simply approximated as 1 in the kinematic region under consideration.   With this effective coupling, one can readily derive the amplitude for the EM production of dipion,
\begin{eqnarray}
{\cal M}_{\gamma \gamma \rightarrow \pi\pi}
 =2e^2 \left [
\epsilon_{\perp1}^\gamma \! \cdot \epsilon_{\perp2}^\gamma  -\frac{2P_\perp^2}{P_\perp^2+m_\pi^2}
(\epsilon_{\perp1}^\gamma \! \cdot \hat P_\perp) (\epsilon_{\perp2}^\gamma\cdot \hat P_\perp)
 \right ]
\end{eqnarray}
Since the incident photons are linearly polarized, the polarization vectors 
$\epsilon_{\perp2}$ and $\epsilon_{\perp1}$ will be replaced with the corresponding photons transverse momenta respectively in the rest of calculations. 
Following the method outlined in Refs.~\cite{Xing:2020hwh}, the impact parameter dependent cross section computed at the lowest order QED reads,
 \begin{eqnarray}
&& \!\!\!\!\!\!\!\!\!\!
\frac{d\sigma_{\gamma \gamma \rightarrow \pi\pi}}{d^2 p_{1\perp} d^2 p_{2\perp} dy_1 dy_2 d^2 \tilde b_\perp }=
\frac{\alpha_e^2}{Q^4}\frac{1}{(2\pi)^2}
  \int d^2k_{\perp} d^2 \Delta_\perp d^2
k_{\perp}' \delta^2( q_\perp-k_{\perp}-\Delta_\perp) e^{i
(k_{\perp}- k_{\perp}') \cdot \tilde b_\perp}
\nonumber \\&&  \times 4 \!\left [\hat k_{\perp}\! \cdot \hat \Delta_\perp  \!-\frac{2P_\perp^2}{P_\perp^2+m_\pi^2}
(\hat k_{\perp} \! \cdot \hat P_\perp) ( \hat \Delta_\perp \cdot \hat P_\perp) \right ]\!\!
\left [\hat k_{\perp}' \! \cdot \hat \Delta_\perp' \! -\frac{2P_\perp^2}{P_\perp^2+m_\pi^2}
(\hat k_{\perp}' \! \cdot \hat P_\perp) ( \hat \Delta_\perp' \cdot \hat P_\perp) \right ]
\nonumber \\&&  \times \
{\cal F}(x_1,k_{\perp}^2){\cal F}^*(x_1,k_{\perp}'^2){\cal F}(x_2,\Delta_\perp^2){\cal F}^*(x_2,\Delta_\perp'^2)
\end{eqnarray}
Here we only consider one photon exchange approximation. The multiple photon re-scattering 
effect is power suppressed in the invariant mass region near the $\rho^0$ resonance peak due to the involvement of  the Weizs$\ddot{a}$cker-Williams type photon TMD  in this process~\cite{Klein:2020jom}. As a comparison, for the dipole type distribution, the photon TMD receives the significant Coulomb correction~\cite{Sun:2020ygb}.  

The interference term arises from the EM amplitude and the photon-nuclear scattering amplitude contributes to the cross section, 
\begin{eqnarray}
 && \!\!\!\!\!\!\!\!\!\!\!\!\!\!\!\!\!\!\!\!\!\!\!\!\!
\frac{d \sigma_I}{d^2 p_{1\perp} d^2 p_{2\perp} dy_1 dy_2 d^2 \tilde b_{\perp} }  =\frac{\alpha_e}{Q^2} \frac{1 }{(2\pi)^4} \frac{1}{\sqrt{4\pi}} \frac{2M_\rho \Gamma_\rho |P_\perp| f_{\rho \pi \pi}}{(Q^2\!-M_\rho^2)^2\!+M_\rho^2 \Gamma_\rho^2}  \int \!\! d^2 \Delta_\perp d^2k_\perp d^2 k_\perp'
\nonumber \\ &&\!\!\!\!\!\!\!\!\!\! \times \delta^2(k_\perp+\Delta_\perp-q_\perp)    \left [\hat k_\perp \! \cdot \hat \Delta_\perp-\frac{2P_\perp^2}{P_\perp^2+m_\pi^2}  (\hat k_\perp \! \cdot \hat P_\perp) (\hat \Delta_\perp \cdot \hat P_\perp) \right ](\hat P_\perp\! \cdot \hat k_\perp' )
\nonumber \\ && \!\!\!\!\!\!\!\!\!\! \times 2
      \left \{  
   \left [  e^{i \tilde b_\perp \cdot (k_\perp'\!\!-k_\perp)}
   {\cal F}(x_1,k_\perp) {\cal F}(x_2,\Delta_\perp)  {\cal F}(x_1,k_\perp'){\cal A}_{co}^*(x_2, \Delta_\perp')     \right ]   \right .\
\nonumber \\  &&+  \left .\
        \left [ e^{i \tilde b_\perp \cdot (\Delta_\perp'\!-k_\perp)}
        {\cal F}(x_2,k_\perp)  {\cal F}(x_1,\Delta_\perp) {\cal F}(x_2,k_\perp') {\cal A}_{co}^*(x_1, \Delta_\perp')     \right ] 
        \right \}\!
\end{eqnarray}
In arriving at the above result, we have assumed that the dipole amplitude is purely imaginary.  It is interesting to notice that due to the presense of angular structures $(\hat k_\perp \! \cdot \hat \Delta_\perp)(\hat P_\perp\! \cdot \hat k_\perp' )$ and $(\hat P_\perp\! \cdot \hat k_\perp' )(\hat k_\perp \! \cdot \hat P_\perp) (\hat \Delta_\perp \cdot \hat P_\perp)$, this interference term vanishes once carrying out the integration over the azimuthal angle of either $q_\perp$ or $P_\perp$. Therefore, it does not contribute to the azimuthal averaged cross section, but instead leads to  $\cos \phi$ and $\cos 3\phi$ azimuthal asymmetries.  The emergencies of these nontrivial azimuthal correlations can be intuitively understood as follows: the linearly polarization state is the superposition of different helicity states. As illustrated in Fig.1, supposing that two incoming photon both carry  spin angular momentum 1 in the amplitude while the incident photon carries spin angular momentum -1 in the conjugate amplitude, the orbital angular momentum carried by the produced dipion system is 2 in the amplitude and -1 in the conjugate amplitude correspondingly.  Such interference amplitude leads to a $\cos 3\phi$ angular modulation. The similar argument applies to the $\cos \phi$ azimuthal asymmetry case.

\begin{figure}[htpb]
\includegraphics[angle=0,scale=0.8]{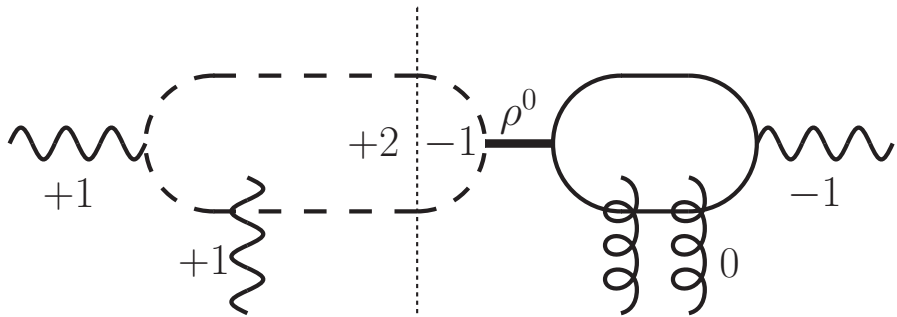}
\caption{An illustration of the mechanism  giving rise to $\cos 3\phi$ azimuthal asymmetry. The solid line represents the quark propagator, while the pion propagator is indicated by the dashed line.   } \label{fig1}
\end{figure}

\section{numerical estimations}
  We now introduce  all ingredients that are necessary for numerical calculations. Most of them follow  our previous work~\cite{Xing:2020hwh}. We first introduce the parametrization for the $b_\perp$ dependent dipole-nucleus and dipole-nucleon scattering amplitudes.  
 For the  dipole-nucleus scattering amplitude, we adopt the parametrization~\cite{Kowalski:2003hm,Kowalski:2006hc},
\begin{eqnarray}
N(b_\perp, r_\perp)=1-e^{-2\pi B_p A T_A(b_\perp) {\cal N}(r_\perp)}
\end{eqnarray}
 Here the nuclear thickness function $T_A(b_\perp)$ is computed with the Woods-Saxon distribution, and $B_p = 4{\text{ GeV}}^2$ in the IPsat model. For the dipole-nucleon  scattering amplitude,  we adopt  a conventional parameterization, namely, the GBW model~\cite{GolecBiernat:1998js,GolecBiernat:1999qd}: ${\cal N}(r_\perp)=1-e^{-\frac{Q_s^2 r_\perp^2}{4}}$.  For the scalar part of  vector meson wave function, we use "Gaus-LC" wave function also taken from Ref.~\cite{Kowalski:2003hm,Kowalski:2006hc}.
\begin{eqnarray}
\Phi^*(|r_\perp|,z)= \beta z(1-z) \exp \left[-\frac{r_\perp^2}{2R_\perp^2}\right ]
\end{eqnarray}
with $\beta=4.47$, $R_\perp^2=21.9 \ \text{GeV}^{-2}$ for $\rho$ meson.
 
 ${\cal F}(x,k_\perp)$ describes the probability amplitude for finding a photon carries certain momentum. The squared  ${\cal F}(x,k_\perp)$  is just the standard   photon TMD distribution $f(x,k_\perp)$. At low transverse momentum it can be commonly computed with  the  equivalent photon  approximation(also often referred to as  the Weizs$\ddot{a}$cker-Williams method) which has been widely used to compute UPC observables(see for example~\cite{Klein:2018fmp,Zha:2018tlq,Klein:2020jom}). In the equivalent photon approximation, $ {\cal F}(x,k_\perp)$    reads,
\begin{eqnarray}
  {\cal F}(x,k_\perp)=\frac{Z \sqrt{\alpha_e}}{\pi} |k_\perp|
 \frac{F(k_\perp^2+x^2M_p^2)}{(k_\perp^2+x^2M_p^2)}
\end{eqnarray}
where $x$ is constrained according to $x= \sqrt{\frac{P_\perp^2+m_\pi^2}{S}}(e^{y_1}+e^{y_2})$. $M_p$ is proton mass. $F$ is the nuclear charge form factor which is assumed to take  the Woods-Saxon distribution as well,
\begin{eqnarray}
F(\vec k^2)= \int d^3 r e^{i\vec k\cdot \vec r} \frac{C^0}{1+\exp{\left [(r-R_{A})/d\right ]}}
\end{eqnarray}
where  $R_{A}$(Au: 6.38 fm, pb: 6.62 fm) is the radius  and d(Au:0.535 fm, Pb:0.546 fm) is the skin depth. $C^0$ is a normalization factor.
\begin{figure}[htpb]
  \includegraphics[angle=0,scale=0.42]{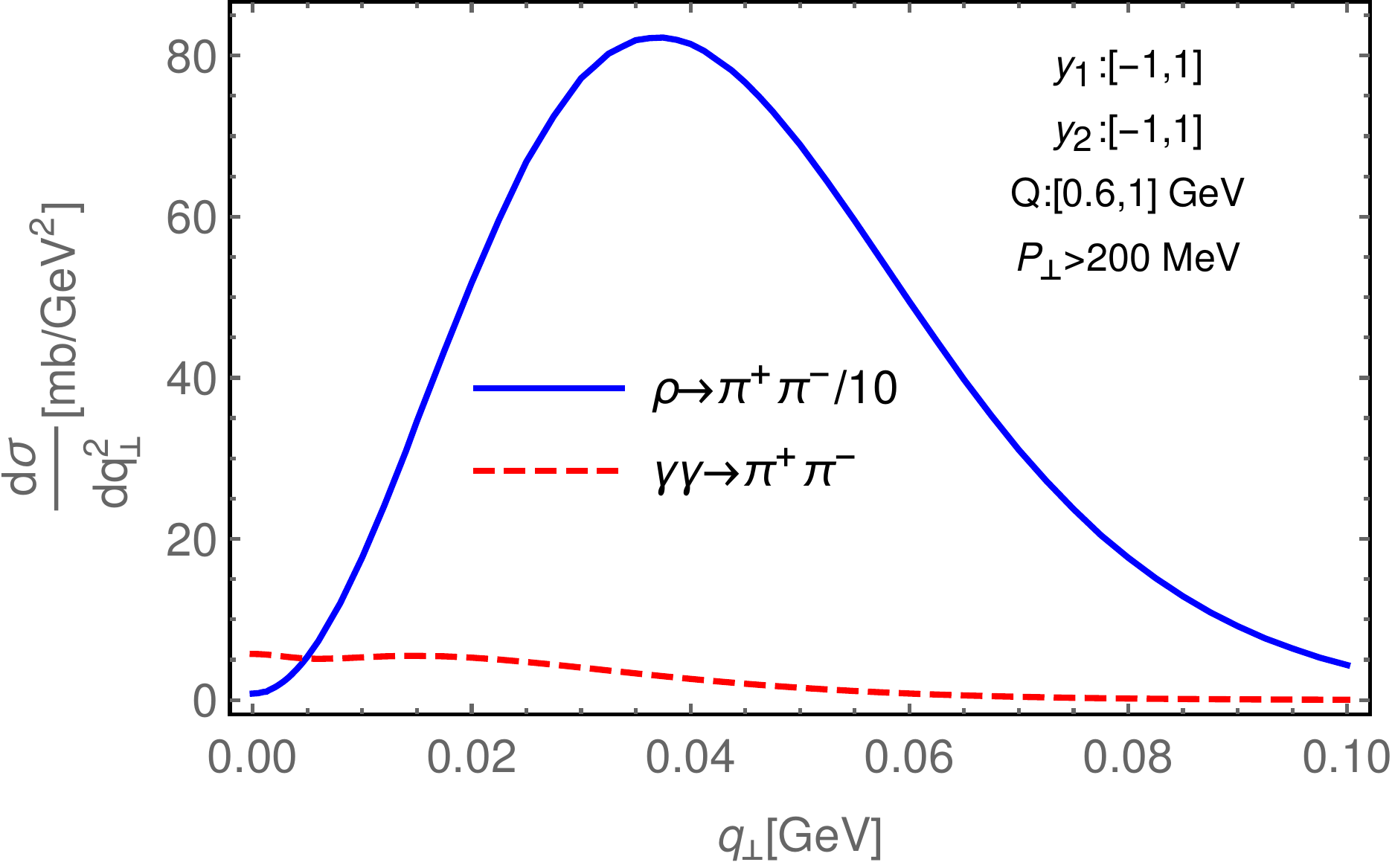}
  \includegraphics[angle=0,scale=0.378]{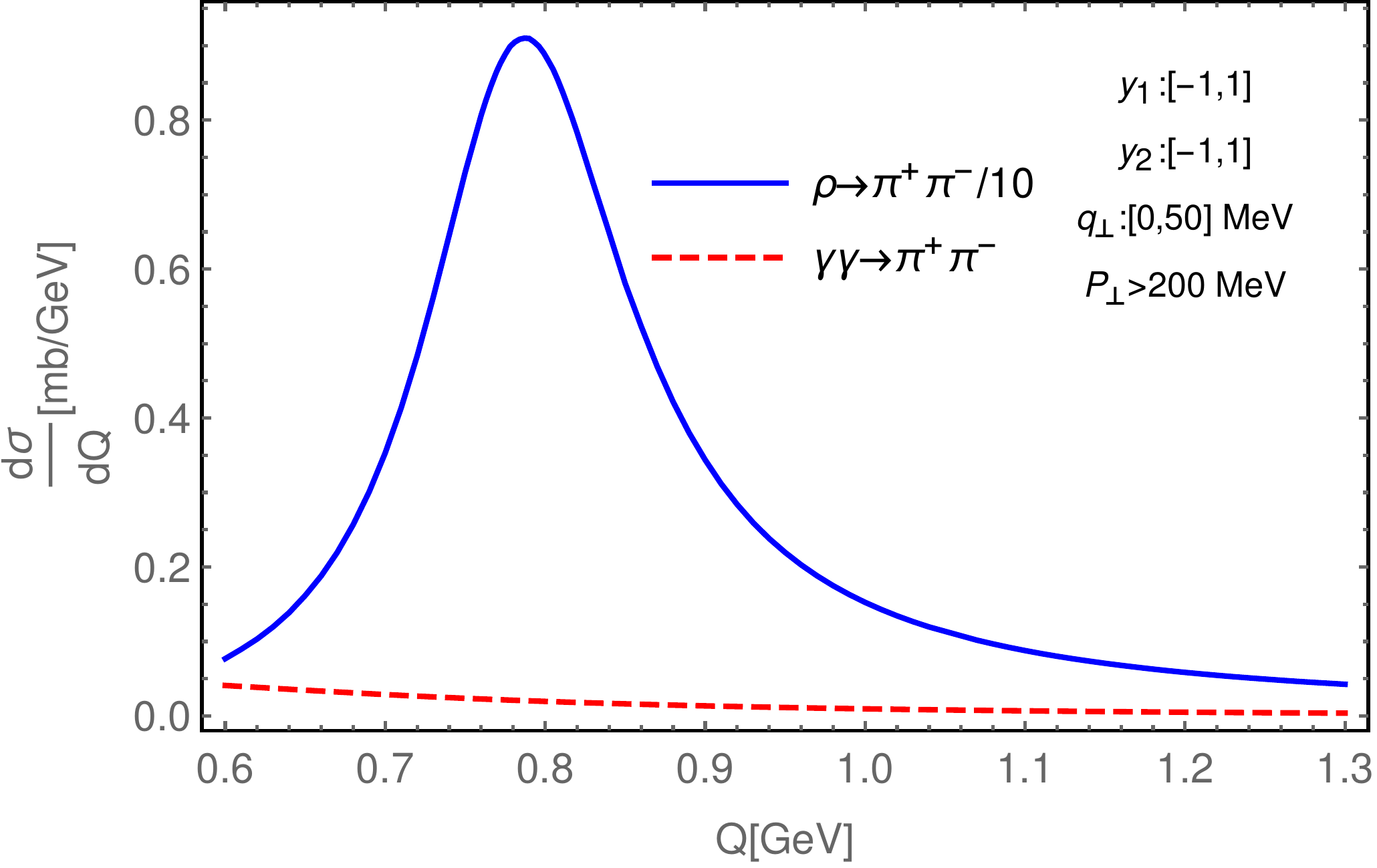}
  \caption{The unpolarized differential  cross section as the function of $q_\perp$(left panel) and the invariant mass(right panel).  The photonuclear  cross section $d \sigma_{\rho \rightarrow \pi \pi}$ is rescaled by a factor 1/10.   } 
  \label{unp}
\end{figure}

UPC events are usually triggered at RHIC and LHC by detecting neutrons emitted at forward angles from a scattered nuclei.  The probability for emitting a neutron is strongly dependent of the impact parameter.  To  account for this effect, one should integrate the differential cross section  over impact parameter range from $2R_A$ to $\infty$ with a weight function, 
\begin{eqnarray}
2 \pi \int_{2R_A}^{\infty} \tilde b_\perp d\tilde b_\perp P^2(\tilde b_\perp) d \sigma(\tilde b_\perp, \ ...)
\end{eqnarray}
 where the probability $P(\tilde b_\perp)$ of emitting a neutron from  the scattered nucleus is conventionally  parameterized as~\cite{Baur:1998ay}
$P(\tilde b_\perp)= P_{1n}(\tilde b_\perp) \exp \left [-P_{1n}(\tilde b_\perp)\right ]$
which is denoted as the ``1'' event, while for emitting any number of neutrons (``X'' event),  the probability is given by $P(\tilde b_\perp)= 1-\exp \left [-P_{1n}(\tilde b_\perp) \right ]$
with $P_{1n}(\tilde b_\perp)= 5.45*10^{-5} \frac{Z^3(A-Z) }{A^{2/3} \tilde b_\perp^2 } \ \text{fm}^2 $. All the subsequent numerical estimations are made for the ``Xn'' event. Having all these parameters specified, we are now ready to perform numerical calculation  of the azimuthal asymmetries for dipion production in UPCs.

We first present the contributions from the EM interaction and the photonuclear reaction to the azimuthal averaged cross section as the function of the invariant mass and pair transverse momentum in Fig.\ref{unp}.  One can see that the EM contribution is not negligible  at low transverse momentum and in the low invariant mass region.  It is worthy to emphasize again that the interference  term drops out in the azimuthal averaged cross section. 
 The suppression of the unpolarized cross section at low $q_\perp<40$ MeV  is due to the destructive interference that was first discussed in Ref.~\cite{Klein:1999gv}.

\begin{figure}[htpb]
  \includegraphics[angle=0,scale=0.38]{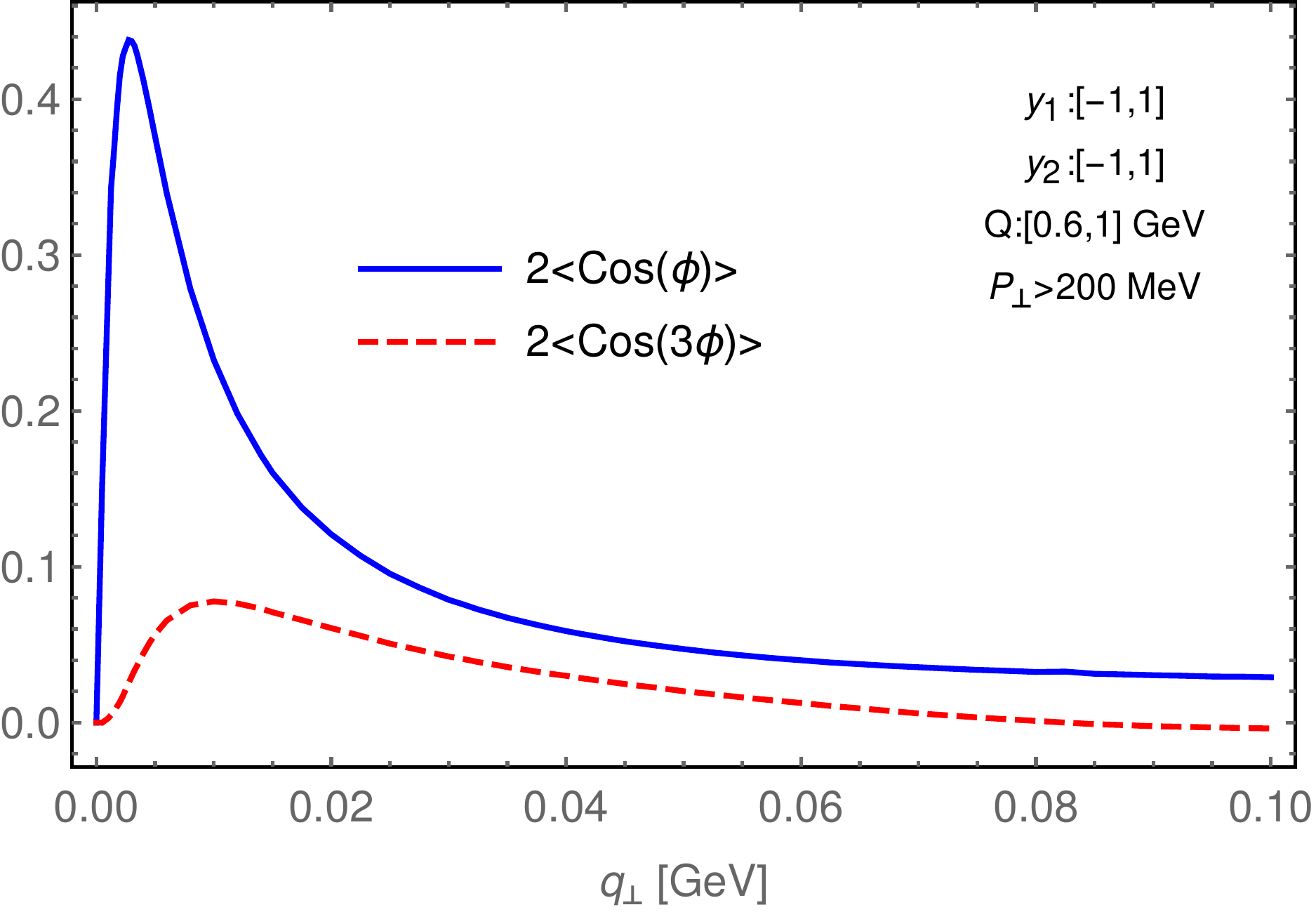}
  \includegraphics[angle=0,scale=0.35]{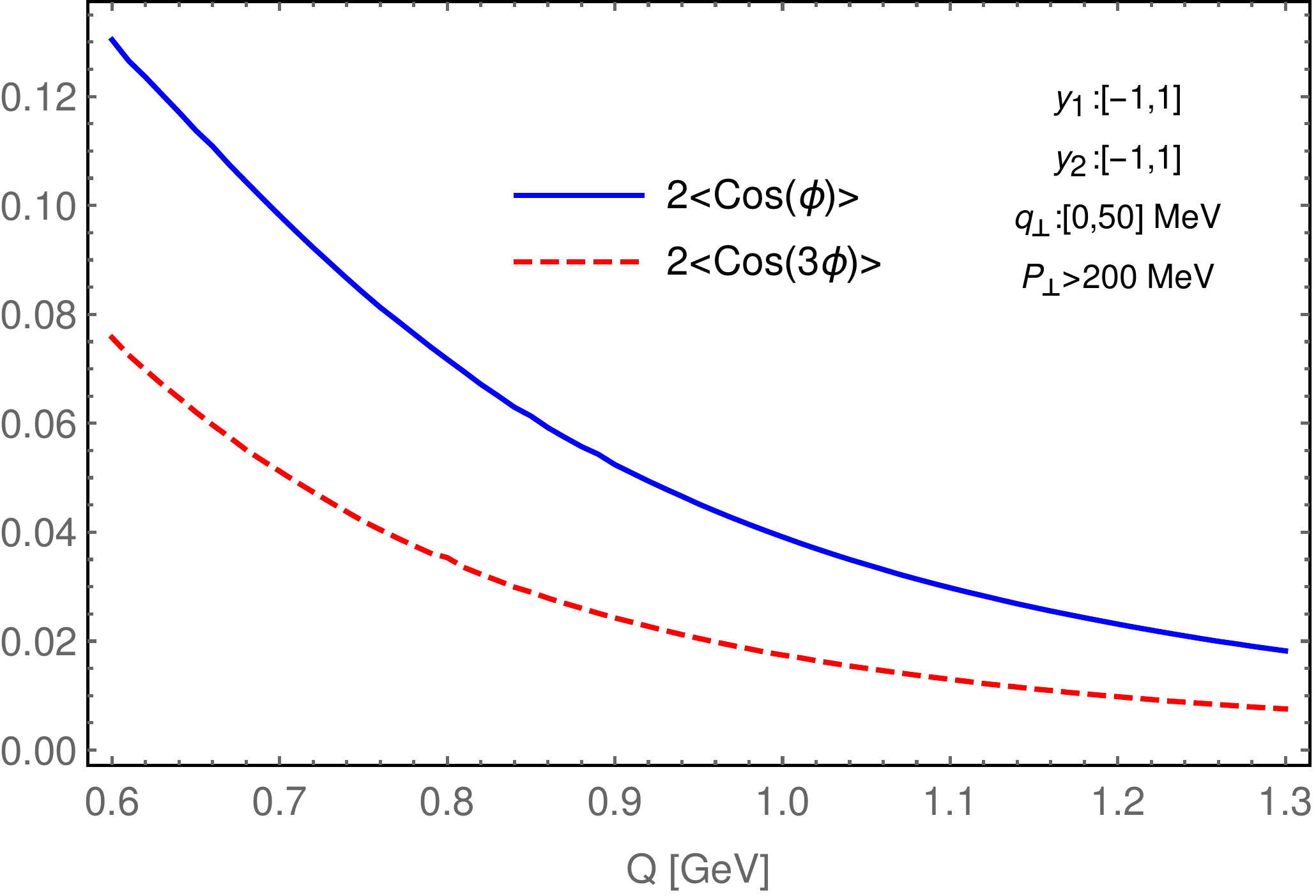}
  \caption{$\cos \phi$ and $\cos 3\phi$ azimuthal asymmetries as the function of $q_\perp$(left panel) and the invariant mass(right panel).    } \label{asy}
\end{figure}

The azimuthal asymmetries from the interference term are displayed in Fig.~\ref{asy}, where the  asymmetries, i.e. the average value of  $\cos 3\phi$ and  $\cos \phi$ are defined as,
\begin{eqnarray}
\langle \cos( n\phi) \rangle =\frac{ \int \frac{d \sigma}{d {\cal P.S.}} \cos n \phi \ d {\cal P.S.} }
{\int \frac{d \sigma}{d {\cal P.S.}}  d {\cal P.S.}}
\end{eqnarray}
with $n=1,3$.  These asymmetries are calculated at the  RHIC energy $\sqrt{S}=200 {\text{ GeV}}$.
Both $\langle \cos \phi \rangle $ and  $\langle \cos 3\phi \rangle $  reach the maximal value at very low $q_\perp$ and then start to decrease with increasing $q_\perp$.  They remain sizable and are roughly few percentage level in the transverse momentum region $20 \ {\text {MeV}} <q_\perp<60 {\text { MeV}}$.   The analysing power is basically of the same order of that from the helicity flip amplitude in elastic proton-proton scatterings, where  the interference at low transverse momentum between the EM and hadronic  contributions has long been used as a tool in the study of the phase of the hadronic amplitude~\cite{Buttimore:1978ry,Buttimore:1998rj} (see also a recent development~\cite{Hagiwara:2020mqb} ).

\

\section{conclusion}
In summary, we have studied the $\cos \phi$ and $\cos 3\phi$ azimuthal angular correlations  in exclusive $\pi^+ \pi^-$  pair production near $\rho^0$ resonance peak in  ultraperipheral heavy ion collisions, where $\phi$ is defined as the angle between pion's transverse  momentum and pair's transverse momentum. The asymmetry essentially results from the linear polarization of incident coherent photons. We focus on low pair transverse momentum  where the electromagnetic and nuclear amplitudes of dipion production are comparable.   In this work, we demonstrate that $\cos \phi$ and $\cos 3 \phi$ azimuthal asymmetries are the characteristic signals of the CNI effect in exclusive pion pair production in UPCs. These azimuthal asymmetries are shown  to be  sizable in the kinematic region of interest, thus can  serve as an alternative method to constrain the phase of the dipole amplitude. It should be feasible  to measure the proposed observables at RHIC and LHC.  We notice that   the direct pion pair continuum receives contributions from several different channels~\cite{Soding:1965nh,Brodsky:1971ud,Klusek-Gawenda:2013rtu,Bolz:2014mya}. Though they unlikely dominate low pair transverse momentum spectrum, it would be interesting to carry out a more comprehensive analysis in a future publication. 

\begin{acknowledgments}
 J. Zhou thanks Zhang-bu Xu,  James Daniel Brandenburg and Chi Yang for helpful discussions.  J. Zhou has been supported by the National Science Foundations of China under Grant No.\ 11675093. Y. Zhou has been supported by the National Science Foundations of China under Grant No.\ 11675092.
\end{acknowledgments}

\bibliography{ref.bib}

\end {document}